\documentclass[aps,prd,nofootinbib]{revtex4}
\usepackage{graphicx}
\usepackage{rotating}

\begin{document}
\title{A ``Baedecker'' for the Dark Matter Annihilation Signal} 
\author{N.W. Evans $^1$, F. Ferrer $^2$ and S. Sarkar $^2$}
\medskip
\affiliation{$^1$ Institute of Astronomy, University of Cambridge,
                  Madingley Road, Cambridge CB3 OHA, UK \\ 
             $^2$ Theoretical Physics, University of Oxford, 
                  1 Keble Road, Oxford OX1 3NP, UK}

\begin{abstract}
We provide a ``Baedecker'' or travel guide to the directions on the
sky where the dark matter annihilation signal may be expected. We
calculate the flux of high energy $\gamma$-rays from annihilation of
neutralino dark matter in the centre of the Milky Way and the three
nearest dwarf spheroidals (Sagittarius, Draco and Canis Major), using
realistic models of the dark matter distribution. Other investigators
have used cusped dark halo profiles (such as the Navarro-Frenk-White)
to claim a significant signal. This ignores the substantial
astrophysical evidence that the Milky Way is not dark-matter dominated
in the inner regions. We show that the annihilation signal from the
Galactic Centre falls by two orders of magnitude on substituting a
cored dark matter density profile for a cusped one. The present and
future generation of high energy $\gamma$-ray detectors, whether
atmospheric Cerenkov telescopes or space missions like GLAST, lack the
sensitivity to detect any of the monochromatic $\gamma$-ray
annihilation lines. The continuum $\gamma$-ray signal above 1 GeV and
above 50 GeV may however be detectable either from the dwarf
spheroidals or from the Milky Way itself. If the density profiles of
the dwarf spheroidals are cusped, then the best prospects are for
detecting Sagittarius and Canis Major. However, if the dwarf
spheroidals have milder, cored profiles, then the annihilation signal
is not detectable. For GLAST, an attractive strategy is to exploit the
wide field of view and observe the Milky Way at medium latitudes, as
suggested by Stoehr {\em et al.} This is reasonably robust against
changes in the density profile.
\end{abstract}

\maketitle

\section{Introduction}

The foremost candidate for the cold dark matter (CDM) composing
galactic haloes is the lightest neutral supersymmetric particle,
namely the neutralino \cite{Jungman}. If so, then neutralino pair
annihilation may lead to observable consequences, in particular the
emission of high energy $\gamma$-radiation \cite{Bergstrom98}. The
possibility that such $\gamma$-rays may be identified by forthcoming
atmospheric Cerenkov telescopes (ACT) such as VERITAS \cite{veritas}
or by satellite-borne detectors like GLAST \cite{glast} has excited
considerable recent interest \cite{baltz,tyler,ts,stoehr}.

It is clearly of importance to identify the best places to search for
such an annihilation signal. Inspired by the highly cusped models
based on numerical simulations of dark halo formation
\cite{navarro,moore}, a number of investigators have suggested that
the centre of the Milky Way may be the optimum target. For example,
Bergstrom {\it et al.} \cite{Bergstrom98} have shown that if the dark
matter density is cusped as $1/r$ at small radii, then the
$\gamma$-ray flux would be detectable for typical neutralino
properties in the minimal supersymmetric extension of the Standard
Model. Inspired also by the persistence of substructure in numerical
simulations, a number of authors \cite{Bergstrom99,ts,crm} have argued
that a substantial enhancement in the $\gamma$-ray signal can be
expected from such `clumps'. In these calculations, the inner regions
of the substructure are also usually assumed to be cusped. However,
even within the framework of the cusped models favored by cosmological
simulations, these conclusions have been contested as being overly
optimistic \cite{stoehr}.

More awkwardly, there is a substantial body of astrophysical evidence
that the halo of the Milky Way is not cusped at all \cite{evansidm}.
First, the microlensing optical depth towards the Galactic Center is
very high. Particle dark matter does not cause microlensing, whereas
faint stars and brown dwarfs do. The total amount of all matter within
the Solar circle is constrained by the rotation curve, so this tells
us that lines of sight towards the Galactic Center are not dominated
by particle dark matter. More specifically, haloes as strongly cusped
as $1/r$, normalised to the local dark matter density as inferred from
the stellar kinematics in the solar neighbourhood, are ruled out by
the high microlensing optical depth \cite{be}. Second, the pattern
speed of the Galactic bar is known to be fast from hydrodynamical
modelling of the motions of neutral and ionised gas. If dark matter
dominates the central regions of the Milky Way, then dynamical
friction will strongly couple the dark matter to the Galactic bar and
cause it to decelerate on a few bar rotation timescales
\cite{debs}. It is now largely accepted by astronomers that bright
galaxies like the Milky Way do not have cusped dark haloes today, with
some investigators suggesting that feedback from star formation may
provide a resolution with cold dark matter theories \cite{bgs}.

In fact, there is {\em no} observational evidence whatsoever that any
nearby galaxy has a cusped dark halo profile. The rotation curves of
low surface brightness and dwarf spiral galaxies have been the subject
of a long controversy \cite{mcgaugh,bosch}. The effects of beam
smearing mean that the H~I rotation curves of many dwarf spirals are
broadly compatible with both cores and cusps. However, the H~II
rotation curves for at least some dwarf spirals are not compatible
with cusps \cite{blais}. Most dwarf spheroidals (dSphs) do not contain
gas and so the structure of the dark halos must be inferred from
stellar motions. Very recently, the survival of kinematically cold
substructure in the Ursa Minor dSph has been used to argue against a
cusped halo \cite{kwge}. Hence, even at the least massive and most
dark matter dominated end of the galaxy mass spectrum, the predictions
of cold dark matter theories concerning halo structure seem to
disagree with the observations. Nonetheless, for the three nearest
dSphs -- Draco, Sagittarius and Canis Major -- there is no direct
evidence either for or against central cusps in their dark matter
distribution.

Given this weight of evidence, it seems very prudent to use both cored
and cusped halo models to estimate the range of the expected
$\gamma$-ray annihilation signal. In this paper, we examine four
possible locations -- the Galaxy Center and the centers of the three
nearest dark-matter dominated dwarf spheroidals (Draco, Sagittarius
and Canis Major). In Section~II we use the most recent data on the
velocity dispersion of the dSphs to constrain a variety of dark halo
models, and in Section~III evaluate the $\gamma$-ray flux from
neutralino self-annihilations. Sections~IV and V summarise the
expected contribution from the background and the criterion for
detection respectively. The results for second generation ACTs and for
GLAST are given in Section~VI.

\section{Models of Dwarf Spheroidals}

Dwarf spheroidals (dSphs) warrant attention because they are amongst
the most extreme dark matter dominated environments. For example, the
mass-to-light ratio of Draco is $\sim 250$ in Solar units \cite{kweg},
while that of the Sagittarius is $\sim 100$ \cite{iwgis}. The recently
discovered possible dSph in Canis Major seems similar to the
Sagittarius in structural properties and dark matter content
\cite{rod}.  Given the seeming absence of dark matter in globular
clusters, dSphs are also the smallest systems dominated by dark
matter.

We develop two sets of models of dSphs. The first set is the cored
spherical power-law models \cite{evans9394}:
\begin{equation}
\rho_{\rm pow}(r) \equiv \frac{v_a^2 r_{\rm c}^\alpha}{4 \pi G} 
 \frac{3 r_{\rm c}^2+r^2 (1-\alpha)}{(r_{\rm c}^2+r^2)^{2+\alpha/2}}.
\label{eq:dmiso}
\end{equation}
Here, $r_{\rm c}$ is the core radius and $v_a$ is a velocity
scale. When $\alpha=0$, the model has an asymptotically flat rotation
curve and is the cored isothermal sphere. The rotation curves of dwarf
galaxies may be gently rising or falling at large radii, so we also
consider models with $\alpha =-0.2$ and $0.2$ respectively.

The second set of models is the cusped haloes
\begin{equation}
\rho_{\rm cusp}(r) \equiv \frac{A}{r^\gamma (r+r_{\rm s})^{3-\gamma}},
\label{eq:dmcusp}
\end{equation}
favored by numerical simulations. Here, $r_{\rm s}$ is the scale
radius and $A$ is the overall normalisation. When $\gamma=1.5$, the
model is the highly cusped Moore {\it et al.} \cite{moore} profile, when
$\gamma =1 $, the model is the Navarro-Frenk-White (NFW) profile
\cite{navarro}.  Additionally, we study the case $\gamma = 0.5$ which
represents a still milder cusped profile.

The two free parameters determining the shape of the profile are set
by fitting to observational data on the Draco dSph using the Jeans
equation \cite{bt}. For a spherical galaxy, the enclosed mass $M(r)$
is related to observables via
\begin{equation}
M(r) = 
 - \frac{r\langle v_r \rangle^2}{G}
 \left(\frac{{\rm d}\log\nu}{{\rm d}\log r} 
 + \frac{{\rm d}\log\langle v_r^2 \rangle}{{\rm d}\log r} 
 + 2\beta\right).
\label{eq:fitgen}
\end{equation}
Here, $\nu$ is the luminosity density, $\langle v_r^2 \rangle$ is the
radial velocity dispersion of the stars and $\beta$ is the anisotropy
of the stellar motions. The luminosity density of Draco $\nu$ is taken
as \cite{kleyna1}:
\begin{equation}
\nu = \frac{\nu_0 r_0^5}{(r_0^2+r^2)^{5/2}},
\label{eq:lum}
\end{equation}
with $r_0=9.71' \approx 0.23$ kpc (using a heliocentric distance for
Draco of $82$~kpc). There are 6 observational points showing the line
of sight velocity dispersion of Draco at different radii in
\cite{kleyna2} (using the data with no rotation subtracted). The
datapoints are consistent with a flat profile between $2'$ and
$22'$. Assuming that the anisotropy now vanishes, then the radial
velocity dispersion is equal to the line-of-sight velocity
dispersion. Finally, the left-hand side of eq.~(\ref{eq:fitgen}) is
fitted to the known right-hand side at the locations of the datapoints
in \cite{kleyna2}, thus giving estimates for the two unknown
parameters for each density profile as quoted in
Table~\ref{tb:profile}.

This algorithm provides models of the Draco dSph that satisfy the
available observational data. Unfortunately, the radial variation of
the velocity dispersion has not been measured for the Sagittarius
dSph. However, the central line-of-sight velocity dispersion of
Sagittarius dSph is 11.4 kms$^{-1}$, very similar to that of Draco (10
kms$^{-1}$). Henceforth, we assume that the underlying structural
parameters ($v_{\rm a}, r_{\rm c}$ or $A, r_{\rm s}$) of the
Sagittarius dSph are the same as Draco. The third dSph under study --
Canis Major -- has only recently been claimed and the evidence for its
existence is not yet clear-cut. There is certainly a surprising
concentration of stars in the direction of Canis Major, but this could
be due to some dynamical feature like an outer spiral arm associated
with the Milky Way disk. However, for the purposes of this study, we
assume the interpretation of the data in terms of a merging dSph is
correct and that Canis Major dSph is similar in size and structure to
Sagittarius.

To determine the extent of the dark matter halo of the dSphs, the
tidal radius must be estimated.  The approximate method used
conventionally is derived from the Roche criterion. The tidal radius
is found by requiring that the average mass in the dSph is equal to
the average interior mass in the Milky Way halo, namely
\begin{equation}
\frac{M_{\rm dSph}(r_{\rm t})}{r_{\rm t}^3} 
 = \frac{M_{\rm MW}(r_{\rm dSph}-r_{\rm t})}{(r_{\rm dSph}-r_{\rm t})^3}.
\label{eq:tidalm}
\end{equation}
Here, $M_{\rm MW} (r)$ and $M_{\rm dSph} (r)$ are the masses enclosed
within radius $r$ of the Milky Way halo and the dwarf spheroidal
respectively, while $r_{\rm dSph}$ is the distance from the Galactic
Center to the centre of the dSph. We remark that this is not the same
as the procedure used in a number of recent papers
\cite{tyler,blasisheth,olinto}, in which the local density at the
tidal radius in the dSph is set equal to the density of the background
Milky Way halo at the center of the dSph.

The results depend on the choice of profile for the Milky Way halo.
For comparison purposes, we consider both a cored isothermal profile
with $r_c=10$ kpc and $v_a=220$ kms$^{-1}$, and a NFW profile with
concentration parameter $c=10$. The total mass of the Milky Way halo
is fixed at $M_{\rm MW} \sim 10^{12}M_\odot$, as suggested in
\cite{we99}. We show in Table~\ref{tb:profile} the results when
eq.~(\ref{eq:tidalm}) is used to determine the tidal radius of a dSph
at the locations of Draco and Sagittarius, for the two adopted models of
the Milky Way halo.

\begin{table}[t]
\begin{center}
Cored Power-Law Models
\end{center}
\begin{tabular}{|c|c|c|c|c|c|} \hline
$\alpha$ & $v_a$ km s$^{-1}$ & $r_{\rm c}$ kpc & $r_{\rm t}$ kpc 
& $r_{\rm t}$ kpc & $M(r_t) \div 10^8 M_\odot$ \\
         &                   &                 & MW - Iso 
& MW - NFW        & \\ \hline \hline
0.2  & 24.7 & 0.25 & 6.2 (2.16) & 1.3 (0.5)  & 4.6 (2.0)   \\ \hline
0    & 22.9 & 0.23 & 7.8 (2.5)  & 1.4 (0.51) & 9.5 (3.0)   \\ \hline
-0.2 & 20.9 & 0.21 & 10.1 (2.8) & 1.6 (0.52) & 22.43 (4.9) \\ \hline
\end{tabular}
\begin{center}
Cusped Models
\end{center}
\begin{tabular}{|c|c|c|c|c|c|} \hline
$\gamma$ & $A \times 10^7 M_\odot$ & $r_{\rm s}$ kpc & $r_{\rm t}$ kpc 
& $r_{\rm t}$ kpc & $M(r_t) \div 10^8 M_\odot$ \\
         &                         &                 & MW - Iso 
& MW - NFW        & \\ \hline \hline
0.5         & 2.3 & 0.32 & 6.6 (2.5)  & 1.5 (0.6)  & 5.5 (3.1) \\ \hline
1 (NFW)     & 3.3 & 0.62 & 7.0 (2.59) & 1.6 (0.57) & 6.6 (3.5) \\ \hline
1.5 (Moore) & 2.9 & 1.   & 6.5 (2.4)  & 1.5 (0.6)  & 5.5 (2.8) \\ \hline
\end{tabular}
\caption{Parameters of the dark matter halo profiles of the Draco
dSph. The last three columns also give (in parentheses) the values at
the location of the Sagittarius dSph. Two values are given for the
tidal radius, according to whether the Milky Way halo is modelled
with an isothermal power-law model or a NFW model. [Notes: (1) the
models sometimes require a slight velocity anisotropy in the very
innermost parts to ensure everywhere physical stresses in the Jeans
equation, (2) the scale radius $r_{\rm s}$ is constrained to lie below
1 kpc].}
\label{tb:profile}
\end{table}

\section{The Gamma-ray flux}

Let the neutralino mass be $m_\chi$ and its self-annihilation
cross-section be $\langle \sigma v \rangle$. Then, the $\gamma$-ray
flux from neutralino annihilation is given by \cite{Bergstrom98}
\begin{equation}
\Phi_\gamma(\psi) = \frac{N_\gamma\langle\sigma v\rangle}{4\pi m_\chi^2} 
 \times \frac{1}{\Delta\Omega} \int_{\Delta\Omega} {\rm d}\Omega
 \int_{\rm los} \rho^2 [r(s)]\ {\rm d}s,
\label{eq:integrand}
\end{equation}
where $\rho$ is the density of the dSph as a function of distance from
its center $r$, which of course depends on the heliocentric distance
$s$. The integration is performed along the line-of-sight to the
target and averaged over the solid angle $\Delta\Omega$ of the
detector. In particular, $N_\gamma = 2$ for the annihilation of two
non-relativistic neutralinos into two photons ($\chi {\bar \chi}
\rightarrow \gamma\gamma$) and $N_\gamma = 1$ for the annihilation
into a photon and a $Z$ boson ($\chi {\bar\chi} \rightarrow
Z\gamma$). The first part of the integrand (\ref{eq:integrand})
depends on the particular particle physics model for neutralino
annihilations. The second part is a line-of-sight integration through
the dark matter density distribution. We discuss each in detail in the
next two subsections.

\begin{table}[t]
\begin{center}
mSUGRA parameters
\end{center}
\begin{tabular}{|c|c|c|c|} \hline
$m_0$ (GeV)&$m_{1/2}$ (GeV)&$\tan \beta$&$|A_0|$ (GeV)\\\hline
10-10000 & 10-10000&1-60&10-10000 \\ \hline
\end{tabular}
\caption{The portion of the mSUGRA parameter space randomly scanned to
generate the models. Here, $m_0$ and $m_{1/2}$ are respectively the
common scalar and gaugino mass at the unification scale, while $A_0$
is the trilinear parameter and $\tan \beta$ is the ratio of the vacuum
expectation values of the two Higgs fields. The $\mu$ term in the
Lagrangian is allowed to have either sign.}
\label{tb:susyps}
\end{table}

\subsection{Particle Physics Model}

To compute ${N_\gamma \langle \sigma v \rangle}/({4 \pi m_\chi^2})$,
we have to select a supersymmetric model. We focus on minimal
supergravity (mSUGRA) models with universal gaugino and scalar masses
and trilinear terms at the unification scale \cite{sugra}. We use the
computer programme SoftSusy \cite{softsusy} to scan the supersymmetric
parameter space (see Table~\ref{tb:susyps}) and generate $10^5$ models
which have consistent electroweak symmetry breaking and grand
unification. The output at the electroweak scale is fed into the
programme DarkSusy \cite{ds} which computes the relic density and
products of the neutralino annihilations. It also checks that a given
model is not ruled out by present accelerator experiments.

A feasible model is one which is permitted by accelerator limits and
which predicts a relic density in the range $0.005 < \Omega_{\rm CDM}
h^2 < 0.2$. This is somewhat broader than the range $0.09 <
\Omega_{\rm CDM} h^2 < 0.13$ determined by fitting the standard
$\Lambda$CDM model to the WMAP data \cite{wmap}. This is done so as to
incorporate the higher values for $\Omega_{\rm CDM} h^2$ found for
consistent alternative CDM models \cite{blan}. The lower limit is set
by requiring the relic particle to provide most of the dark matter in
galaxies (taking a typical mass-to-light ratio for galaxies of $\sim
10$, cf. the critical mass to light ratio of $\sim 2000$ in solar
units). The range of the supersymmetric parameters scanned are given
in Table~\ref{tb:susyps}. For each feasible model, we record the
quantities $N_\gamma \langle \sigma v \rangle$ for the discrete lines
$\chi \chi \rightarrow \gamma \gamma$ and $\chi \chi \rightarrow Z
\gamma$, as well as the continuous $\gamma$-ray spectrum above 1 and
50 GeV. Note that in the previous study \cite{Bergstrom98}, the
relic density was much less constrained and models with an arbitrary
low values were permitted. 

\subsection{Line-of-Sight Integration of Dark Matter Density}

The line-of-sight integration can be manipulated thus:
\begin{equation}
\langle J \rangle_{\Delta \Omega} \doteq \frac{1}{\Delta \Omega}
\int_{\Delta \Omega} J(\psi) {\rm d}\Omega = \frac{2\pi}{\Delta\Omega}
\int_0^{\theta_{\rm max}} {\rm d}\theta\,\sin\theta 
\int_{s_{\rm min}}^{s_{\rm max}} 
{\rm d}s\,\rho^2 \left(\sqrt{s^2+s_0^2-2 s s_0 \cos\theta}\right)
\label{eq:jav}
\end{equation}
where 
\begin{equation}
J(\psi) = \int_{\rm los} {\rm d}s\,\rho^2(r).
\end{equation}
In these formulae, angled brackets denote the averaging over the solid
angle $\Delta \Omega$, while $s_{\rm min}$ and $s_{\rm max}$ are the
lower and upper limits of the line-of-sight integration, given by $s_0
\cos\theta \pm \sqrt{r_t^2 - s_0^2 \sin^2\theta}$. Here, $s_0$ is the
heliocentric distance of the dSph and $r_t$ is the tidal radius of the
dSph. Finally, $\theta_{\rm max}$ is the angle over which we average
around the center of the dSph. It generally is, at least, equal to the
experimental resolution and can be fixed using:
\begin{displaymath}
\Delta \Omega =2 \pi \int_0^{\theta_{\rm max}} {\rm d}\theta\, \sin\theta 
= 2 \pi (1-\cos(\theta_{\rm max}) ).
\end{displaymath}
The quoted point spread function widths for the various experiments
are: $0.4^\circ$ (EGRET), $0.1^\circ$ (GLAST, HESS and VERITAS),
$0.15^\circ-0.04^\circ$ (CANGAROO-III). EGRET and GLAST are satellite
detectors with low energy thresholds ($\approx 100$ MeV), high energy
resolution ($\approx 15\%$) but only moderate angular precision. The
others are ACTs with higher thresholds ($\approx$ 100 GeV) but better
angular resolution.  Typical reference sizes for the solid angle are
$\Delta \Omega=10^{-5}$ sr for ACTs and GLAST and $\Delta
\Omega=10^{-3}$ sr for EGRET.

\begin{table}[t]
\begin{center}
Cored Power-Law Models
\end{center}
\begin{tabular}{|c||c|c||c|c||c|c|} \hline
$\alpha$ & \multicolumn{2}{c|}{Sagittarius} &
\multicolumn{2}{c|}{Draco} & \multicolumn{2}{c|}{Canis} 
\\
\cline{2-7} & $\Delta \Omega=10^{-3}$ sr & $\Delta \Omega=10^{-5}$ sr
& $\Delta \Omega=10^{-3}$ sr & $\Delta \Omega=10^{-5}$ sr
& $\Delta \Omega=10^{-3}$ sr & $\Delta \Omega=10^{-5}$ sr \\ \hline 
0.2 & 0.6 & 3.4 & 0.07 & 2.2 & 2.4 & 3.4 \\ \hline
0 & 0.6 & 3.3 & 0.06 & 2.2 & 2.4 & 3.5 \\ \hline
-0.2 & 0.6 & 3.2 & 0.07 & 2.2 & 2.4 & 3.4 \\ \hline 
\end{tabular}
\begin{center}
Cusped Models
\end{center}
\begin{tabular}{|c||c|c||c|c||c|c|} \hline
$\gamma$ & \multicolumn{2}{c|}{Sagittarius} &
\multicolumn{2}{c|}{Draco} & \multicolumn{2}{c|}{Canis} \\
\cline{2-7} & $\Delta \Omega=10^{-3}$ sr & $\Delta \Omega=10^{-5}$ sr
& $\Delta \Omega=10^{-3}$ sr  & $\Delta \Omega=10^{-5}$ sr 
& $\Delta \Omega=10^{-3}$ sr  & $\Delta \Omega=10^{-5}$ sr \\ \hline 
0.5 & 1.1 & 17.8  & 0.1 & 5.7 & 6.2 & 32.3\\ \hline
1 (NFW) & 1.3 & 36.9 & 0.1 & 7.2 & 8.3 & 139.9 \\ \hline
1.5 (Moore) & 7.3 & 615.1 & 0.6 & 55.4 & 49.1 &5469  \\ \hline
\end{tabular}
\begin{center}
Galactic Center
\end{center}
\begin{tabular}{|c||c|c|} \hline
Profile & $\Delta \Omega=10^{-3}$ sr & $\Delta \Omega=10^{-5}$ sr \\ \hline 
NFW, $\gamma=1$ & 26 & 280 \\ \hline
Cored, $\alpha=0$ & 0.3 & 0.3 \\ \hline
\end{tabular}
\caption{Values of $\langle J \rangle_{\Delta \Omega}$ for the
Sagittarius, Draco and Canis Major dSphs in units of $10^{23}$
GeV$^2$cm$^{-5}$.  The tidal radius of the dSphs is calculated
assuming an isothermal profile for the Galactic halo. Additionally,
results in the direction of the Galactic Center are given for both the
NFW and isothermal models of the Galactic halo.}
\label{tb:losint}
\end{table}

Table~\ref{tb:losint} lists values of $\langle J \rangle_{\Delta
\Omega}$ for the dSph profiles introduced in Section II. The
heliocentric distances to the Draco, Sagittarius and Canis Major dSphs
are $\sim 80, 24$ and 8 kpc respectively and this largely controls the
relative values of $\langle J \rangle_{\Delta \Omega}$ for the
three dSphs. Clearly, the comparative closeness of the Canis Major
dSph works to its advantage as a possible target. An increase in
angular sensitivity enhances the signal for all three dSphs. We remark
that, in the literature, there is a considerable spread in the values
obtained for $\langle J \rangle_{\Delta \Omega}$ for different
sources. In \cite{baltz}, a 1-component King profile was used to model
the dSph density distribution. These authors only give explicit
estimates of the entire line-of-sight integral. However, the values
are of the order of $10^{21}$GeV$^2$cm$^{-5}$, lower than those
implied by Table~\ref{tb:losint}. In \cite{tyler}, no angular average
is taken, but instead the approximation
\begin{equation}
J \approx \frac{\int_{\rm los} \rho^2(l) {\rm d}l}{4 \pi d^2},
\label{eq:approxj}
\end{equation} 
is used, together with an singular isothermal profile for
Draco. Taking the value given for $m_\chi=100$ GeV implies $3.7 \times
10^{20}$ GeV$^2$ cm$^{-5}$ for the line-of-sight integral. It is
surprisingly low and yet the plots manage to exclude as large a region
in parameter space as in \cite{Bergstrom98}! As we discuss in Section
V, the reason for this is the criterion used in \cite{tyler} to
identify a detectable signal.

As seen in the bottom panel of Table~\ref{tb:losint}, the $\gamma$-ray
emission towards the Galactic Center falls by at least two orders of
magnitude on moving from a cusped NFW halo to a cored isothermal
model. The point that the signal from the Galactic Center depends
sensitively on the assumed halo profile, and so may have been
overestimated, has also been made recently by Stoehr et
al. \cite{stoehr}. For example, an optimistic result for the
$\gamma$-ray flux towards the Galactic Center was obtained in
\cite{Bergstrom98} by using a cusped NFW model normalised to satisfy
two constraints on the halo mass $M$ and circular speed $v_{\rm h}$,
namely
\begin{equation}
M (r < 100\, {\rm kpc}) = (6.3 \pm 2.5) \times 10^{11} M_{\odot},
\qquad\qquad v_{\rm h}(R_0) \approx 128-207\, {\rm km\,s}^{-1}
\end{equation}
If the normalisation is set to obtain the maximum flux (as is done for
example in Fig. 9 of \cite{Bergstrom98}), then the models possess a
local dark matter density substantially in excess of the usual value
of $\sim 0.3$ GeV cm$^{-3}$.  Anyhow, even accepting the debatable
proposition that the Galaxy did once have a pristine dark halo of the
NFW form, the formation of the Galactic disk, bulge and bar will have
substantially re-processed the dark matter distribution. Certainly,
the evidence from the microlensing optical depth towards the Galactic
Centre and the pattern speed of the Galactic bar are inconsistent with
models dominated by dark matter in the central regions
\cite{evansidm}.

Finally, let us illustrate how to convert the numbers in
Table~\ref{tb:losint} into photon fluxes. This requires adopting
characteristic values for the particle physics parameters; here, we
take $m_\chi = 100$ GeV, and $N_\gamma \langle \sigma v \rangle =
10^{-25}$ cm$^3$ s$^{-1}$ for energies in excess of 1 GeV. Assuming
the field of view is fixed at $3^\circ$ (in other words, the
semi-angle of the cone whose axis points towards the center of the
objects is $1.5^\circ$), the implied photon fluxes from the
Sagittarius, Draco and Canis Major dSphs are as given in
Table~\ref{tb:photoncounts}.

\begin{table}[t]
\begin{center}

Typical Expected Flux Values (cm$^{-2}$ s$^{-1}$) for $E >$ 1 GeV
\end{center}
\begin{tabular}{|c||c|c|c|c|} \hline
Model & Galactic Center & Sagittarius dSph & Draco dSph & Canis Major
dSph\\ \hline
NFW       & $3.2 \times 10^{-9}$ & $1.1 \times 10^{-10}$ & $9.9 \times
10^{-12}$ & $7.8 \times 10^{-10}$\\ \hline
Power-law & $6.0 \times 10^{-11}$ & $ 5.3 \times 10^{-11}$ &
$5.1 \times 10^{-12}$ & $2.9 \times 10^{-10}$\\ \hline
\end{tabular}
\caption{We show the annihilation $\gamma$-ray fluxes (cm$^{-2}$ s$^{-1}$) 
for continuum emission above 1 GeV from the Galactic Centre and the three
dSphs, assuming a field of view of $3^\circ$. These numbers are computed
from the values of $\langle J \rangle$ given in Table~\ref{tb:losint}
under further assumptions of characteristic values for the neutralino mass
and cross-section (viz. $m_\chi = 100$ GeV and $N_\gamma \langle \sigma v
\rangle = 10^{-25}$ cm$^3$ s$^{-1}$). In each case, results for a cusped
NFW ($\gamma = 1$) and a cored power-law ($\alpha = 0$) model are given.}
\label{tb:photoncounts}
\end{table}

\section{Computation of the background}

There are three sources of background for the signal under
consideration: hadronic, cosmic-ray electrons and diffuse
$\gamma$-rays from astrophysical processes.  The last is negligible
for ACTs, but is the only one present for satellite experiments like
GLAST or EGRET. Let us consider each source of background in turn.

\subsection{Hadronic and Electronic}

Bergstrom et. al. \cite{Bergstrom98} use data taken with the Whipple
ACT to derive the following expression for the hadronic background:
\begin{equation}
\frac{{\rm d}\Phi_{\rm had}}{{\rm d}\Omega} \left( E > E_0 \right) 
= 6.1 \times 10^{-3}
 \epsilon_{\rm had} \left( \frac{E_0}{1\;{\rm GeV}} \right)^{-1.7}
 {\rm cm}^{-2}{\rm s}^{-1}{\rm sr}^{-1},
\label{eq:berghad}
\end{equation}
where $\epsilon_{\rm had}$ is intended to take into account improved
hadronic rejection expected in future ACTs, but is at present set to
unity.

Showers initiated by cosmic-ray electrons are indistinguishable from
gamma rays. This contribution to the background for ACTs is, according
to \cite{Bergstrom98} (who cite \cite{longair} for this purpose):
\begin{equation}
\frac{{\rm d}\Phi_{e^-}}{{\rm d}\Omega} 
= 3 \times 10^{-2} \left(\frac{E_0}{1\;{\rm GeV}}\right)^{-2.3} 
  {\rm cm}^{-2}{\rm s}^{-1}{\rm sr}^{-1}.
\label{eq:bergelec}
\end{equation}

\subsection{Diffuse Emission}

The diffuse $\gamma$-ray background is usually taken to be dominated
by the Galactic \cite{hunter} or extragalactic \cite{sreekumar}
contribution, depending on whether the target location is the Galactic
Center or at higher latitudes ($b \geq 10^\circ$). For example, a fit
to the EGRET data \cite{hunter} at 1 GeV (dominated by the Galactic
contribution) is given in \cite{Bergstrom98} as
\begin{equation}
\frac{{\rm d}\Phi_{\rm diff}}{d\Omega {\rm d}E}=N_0(l,b) 10^{-6} \left(
\frac{E_0}{1\;{\rm GeV}} \right)^{-2.7}{\rm cm}^{-2}{\rm s}^{-1}{\rm
sr}^{-1}{\rm GeV}^{-1},
\label{eq:bergdif}
\end{equation}
where $N_0(l,b)$ is a factor in the range 1--100, with higher values
for the central regions of the Galaxy. In \cite{baltz}, only the
extragalactic contribution from EGRET, estimated in \cite{sreekumar},
is considered:
\begin{eqnarray}
\frac{{\rm d}\Phi_{\rm diff}}{{\rm d}\Omega dE}&=&\left(7.32 \pm 0.34 \right)
\times 10^{-9} \left(\frac{E_0}{451\; {\rm MeV}} \right)^{-2.10 \pm 0.03}
{\rm cm}^{-2}{\rm s}^{-1}{\rm sr}^{-1} {\rm MeV}^{-1}
 \nonumber\\
&\approx& 1.4 \times 10^{-6} \left( \frac{E_0}{1\;{\rm GeV}} \right)^{-2.10
\pm 0.03}{\rm cm}^{-2}{\rm s}^{-1}{\rm sr}^{-1}{\rm GeV}^{-1}.
\label{eq:baltzdif} 
\end{eqnarray}
So, the spectral indices of the Galactic and extragalactic
contributions are about -2.7 and -2.1 respectively.

However, the separation between the Galactic and extragalactic
background is not clear. For example in \cite{keshet}, the case is
made for a very low extragalactic background. Studying the region
around the Galactic poles ($b\approx90^\circ$), it seems that, even
there, most of the contribution is of Galactic origin. In particular,
the main contribution is not isotropic but correlated with known
Galactic tracers. The EGRET collaboration concede that {\em any}
simple model for the diffuse background is unlikely to work for all
points in the sky and at all energies \cite{hartman}.

To be conservative, we normalize the flux to the EGRET data above
1 GeV and choose a spectral index of -2.1 which is the worst case:
\begin{equation}
\frac{{\rm d}\Phi_{\rm diff}}{{\rm d}\Omega dE} = {\cal N} \left(
\frac{E}{1\, {\rm GeV}}\right)^{-2.1}.
\end{equation}
The emission above 1 GeV in the region of our interest can be
downloaded from the EGRET website \cite{egret}.  The exact values for
the diffuse emission are $6.7 \times 10^{-7} {\rm cm}^{-2}{\rm
s}^{-1}{\rm sr}^{-1}$ at the location of the Draco dSph
($l=86.4^\circ,\;b=34.7^\circ$), $3.18 \times 10^{-6}{\rm cm}^{-2}{\rm
s}^{-1}{\rm sr}^{-1}$ at the location of the Sagittarius dSph
($l=5.6^\circ,\;b=-14.1^\circ$), and $1.2 \times 10^{-4}{\rm
cm}^{-2}{\rm s}^{-1}{\rm sr}^{-1}$ at the Galactic Center.

The diffuse emission is the only background for satellite
experiments. Its large variation with Galactic coordinates can make a
weak source in Draco relatively brighter than strong emission from the
Galactic Center, overwhelming the numbers in
Table~\ref{tb:losint}. For ACTs, however, the hadronic and electronic
backgrounds are much larger and independent of Galactic coordinates,
so the hierarchy from Table~\ref{tb:losint} is retained. So, this
raises the possibility that the Sagittarius dSph might have a higher
signal-to-noise ratio with ACTs, but the Draco dSph is more clearly
seen from satellites.

\section{The detectors}

\subsection{Minimum Detectable Flux}

For the dSphs, the minimum detectable flux $\Phi_\gamma$ is determined
using the prescription that, for an exposure of $t$ seconds made with
an instrument of effective area $A_{\rm eff}$ and angular acceptance
$\Delta \Omega$, the significance of the detection must exceed $5
\sigma$:
\begin{eqnarray}
\frac{\Phi_\gamma 
\sqrt{\Delta\Omega A_{\rm eff} t}}{\sqrt{\Phi_\gamma+\Phi_{\rm bg}}} \ge 5.
\label{eq:numphotons}
\end{eqnarray}
Here, $\Phi_\gamma$ denotes the neutralino annihilation flux in ${\rm
cm}^{-2}{\rm s}^{-1}{\rm sr}^{-1}$, while $\Phi_{\rm bg}$ is the
background flux. Any detector sees some photons from both dark matter
annihilation and background, so the error in the measurement is
$\propto \sqrt{\Phi_\gamma+\Phi_{\rm bg}}$ and not $\propto
\sqrt{\Phi_\gamma}$. As pointed out in \cite{lima}, the use of the
latter formula {\em overestimates} the significance of any detection.

When studying the signal from discrete $\gamma$-lines, $d\Phi_{\rm
bg}/d\Omega$ is the background flux falling under the annihilation
line. If the background has a differential spectrum $d^2\Phi_{\rm
bg}/d\Omega dE=N_0 E^{-\delta}$, and if the energy resolution of the
instrument is $\sigma_E/E$, then the background under a line at energy
$E_0$ (i.e., in the interval $\left[E_0-\sigma_E,E_0+\sigma_E \right]$
containing $68\%$ of the signal) is given by \cite{Bergstrom98}:
\begin{equation}
\frac{{\rm d}\Phi_{\rm bg}}{{\rm d}\Omega} 
= \frac{N_0}{\delta-1} E_0^{-\delta+1} \times 
  \eta \left(\sigma_E/E,\delta \right),
\label{eq:backline}
\end{equation}
with 
\begin{equation}
\eta \left(\sigma_E/E,\delta\right) 
= \frac{1}{\left(1-\sigma_E/E\right)^{1-\delta}} 
  - \frac{1}{\left(1+\sigma_E/E\right)^{1-\delta}}.
\label{eq:eta}
\end{equation}
For ACTs, the background is the sum of three different power laws; for
satellites, only the diffuse background is needed.

In the literature, a number of different algorithms are used to define
a detection of dark matter annihilation. Some authors additionally
require a minimum number of detected photons, though this number is
set somewhat arbitrarily. In \cite{Bergstrom98}, the minimum number is
25 for ACTs and 10 for satellite experiments; in \cite{aharonian}, it
is 100 for ACTs. In \cite{baltz}, no minimum number of detected
photons is required, which allows the possibility of a high
significance detection with a tiny number of received photons.  In
\cite{tyler}, a completely different strategy altogether is used:
constraints on supersymmetric parameter space are found by requiring
that Draco's flux be less than the least significant detection (the
Large Magellanic Cloud at $4\sigma$ \cite{lamb}) above 1 GeV,
resulting in a minimum flux for detection of $10^{-8} {\rm cm}^{-2}
{\rm s}^{-1}$. In this way, the noise enters {\em linearly} into the
expression. This explains why Tyler \cite{tyler} excludes a large
region in mSUGRA parameter space from the non-detection of dSphs,
despite the fact that the values of the integral (\ref{eq:jav})
towards the dSphs are quite low.

\begin{sidewaystable}[ht]
\begin{center}
High energy $\gamma$-ray detectors
\end{center}
\begin{tabular}{|c|c|c|c|c|c|} \hline
& HESS (I) & VERITAS & MAGIC & EGRET & GLAST \\ \hline \hline
Energy & 40 GeV-10 TeV & 50 GeV-10 TeV & 30 GeV-10 TeV & 
20 MeV-30 GeV & 20 MeV-300 GeV \\ \hline 
$\sigma_E/E$ & $\approx 10\%$ & $\approx 15\%$ & $\approx 20\%$ & $< 10\%$ 
& $\approx 5\%>10$ GeV \\ \hline 
$A_{\rm eff}$ (${\rm cm}^2$) & $4 \times 10^8 (>100$ GeV) & $4 \times 10^8 
(>100$ GeV) & $4 \times 10^8 (>100$ GeV) & $1.5 \times 10^3$ & $10^4$ 
\\ \hline 
$\Phi_{\rm min}$ (${\rm cm}^{-2}{\rm s}^{-1}$) & $8 \times 10^{-12}
(>100$ GeV) 
& $9 \times 10^{-12} (>100$ GeV)  & $\approx 10^{-11} (> 100$ GeV)  & 
$10^{-7} (>100$ MeV) & $3 \times 10^{-9} (>100$ MeV) \\  \hline 
Ang. res. (single $\gamma$) & $<0.1^\circ$ at 100 GeV & $<0.1^\circ$ at 100
GeV & $\approx 0.2^\circ$ & $<5.8^\circ$ at 100 MeV & 
\begin{tabular}{c} $2^\circ$ at 100 MeV\\$0.1^\circ$ at 10 GeV \end{tabular} 
\\ \hline 
Field of view & $4.3^\circ-5^\circ$ & $3.5^\circ$ & $\approx 5^\circ$
& 0.5 sr & 2.4 sr \\  \hline \hline 
\end{tabular}
\caption{Performance of the gamma-ray detectors. Numbers quoted
correspond to $5\sigma$ sensitivity after 100 hours of observation for
ACTs and 1 yr for GLAST.}
\label{tb:detectors}
\end{sidewaystable}

\subsection{Performance of the Detectors}

The detector characteristics of the different experiments are
summarised in Table~\ref{tb:detectors}.  For definiteness, we use
$\Delta \Omega=10^{-5}{\rm sr} \approx 0.1^\circ$ for the angular
average when considering ACTs (appropriate for energies $\sim 100$
GeV) or GLAST (10 GeV), and $\Delta \Omega=10^{-3}{\rm sr} \approx
1^\circ$ for EGRET (10 GeV).  Also important is the observation time,
which is chosen as $t \approx 1$~yr for satellites.  For the next
generation ACTs, assuming four telescopes, we use an observation time
$t \approx 100$~h and an exposure $A_{\rm eff}=4 \times 10^8$
cm$^2$. This seems reasonable, as CANGAROO \cite{cangaroo} and the
last phase of HESS \cite{hess} will have four telescopes, while
VERITAS will have as many as seven \cite{veritas}. MAGIC \cite{magic}
uses a single 17~m mirror and has roughly the same performance as
next-generation ACTs, but with a reduced threshold of 30 GeV.

\section{Results}

The following plots show the parts of the supersymmetric parameter
space that can be probed through the detection of a $\gamma$-ray
signal from neutralino annihilations. We typically show the region to
which GLAST and a generic second generation ACT will be sensitive. The
plots found in \cite{Bergstrom98} show the $\gamma$-ray flux in
cm$^{-2}$s$^{-1}$ against photon energy. They are not appropriate for
depicting the exclusion limits from observations of different parts of
the sky because the flux changes and hence so do the points
representing theoretical models. We prefer to use the type of plot
presented in \cite{baltz} with $N_\gamma \langle\sigma v\rangle$
(which depends exclusively on the particle physics model) versus
$m_\chi$ (although other quantities could be used as well).
 
From eq.~(\ref{eq:numphotons}), we write the condition for detection
in a more convenient way for the plots:
\begin{eqnarray}
N_\gamma \langle \sigma v \rangle \ge \frac{4 \pi m_\chi^2}{<J>} 
\frac{25+\sqrt{625+100 \Delta \Omega A_{\rm eff} t \Phi_{\rm bg}}}{2 \Delta \Omega A_{\rm eff} t}
\label{eq:plot}
\end{eqnarray}
Here, we see that increasing the angular acceptance $\Delta \Omega$
can increase the signal to noise ratio. In fact, if both signal and
background are constant, the significance increases (and the minimum
value of $N_\gamma \langle \sigma v \rangle$ that can be probed
decreases) as $\sqrt{\Delta \Omega}$.  However, the signal is not
constant as Table~\ref{tb:losint} shows, and the angular acceptance
that maximizes the significance does not necessarily coincide with the
minimum angular resolution of the detector. So the optimal strategy is
to scan between the minimum angular resolution and the maximum field
of view, choosing the field for which the signal to noise ratio is
maximised.  This depends on the position in the sky and on the type of
profile. For instance, the distant Draco looks like a point source and
the maximum signal is for the smallest angle possible. For
Sagittarius, the optimal angle is $0.4^\circ$ for cored profiles and
the smallest possible for cusped profiles.  In order not to put too
many lines in the plot, we have avoided drawing all the halo types and
show only the extreme cases.

\subsection{Discrete Lines}

\begin{figure}[t]
\begin{center}
\includegraphics[height=9cm]{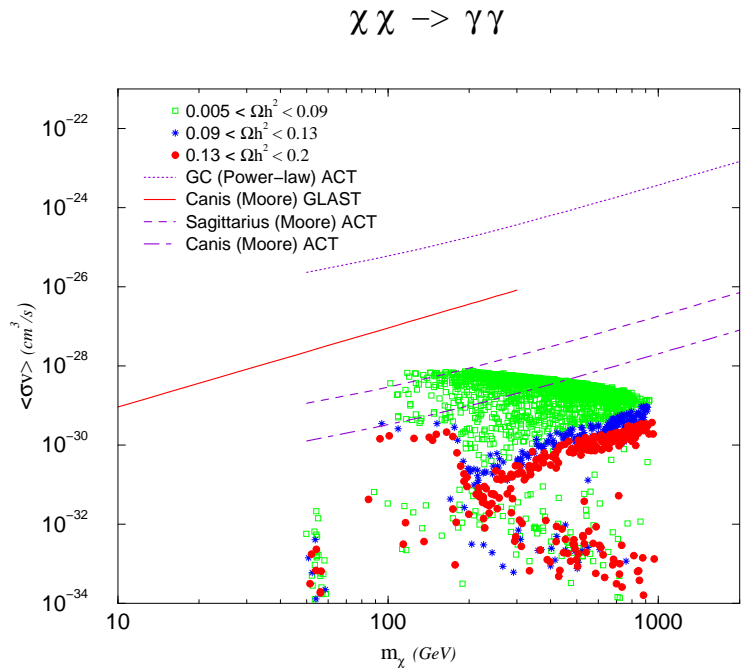}
\end{center}
\caption{Exclusion limits for the discrete line $\chi \chi \rightarrow
\gamma \gamma$. For all the experiments, only the most favorable cases
are shown. The green, red and blue points correspond to mSUGRA models
with $\Omega_{\rm CDM} h^2$ in the range 0.005--0.2, as discussed in
Section IIIA. The red points satisfy the more stringent WMAP
constraints $0.09 < \Omega_{\rm CDM} h^2 < 0.13$.  The exclusion
limits for $\chi \chi \rightarrow Z\gamma$ are very similar and not
shown here.}
\label{fig:twogamma}
\end{figure}
The annihilation of two neutralinos gives rise to two photons with
energy $E_\gamma \approx m_\chi$. The region probed by the different
experiments is shown in Fig.~\ref{fig:twogamma}. Also shown are $\sim
1500$ points in the mSUGRA parameter space that comply with all the
accelerator limits (including $b \rightarrow s \gamma$, $(g-2)_\mu$
and other accelerator limits \cite{pdg} that are incorporated in
DarkSusy). All the points, bar five, have spin-independent
cross-section with protons or neutrons below $10^{-6}$ pb, thus
compatible with limits set by the Edelweiss nuclear recoil detector,
but not the disputed signal claimed by the DAMA experiment
\cite{dmexpt}. Also, the upward-going, neutrino-originated, muon
showers have a flux of $\leq 10^{4}$ km$^{-2}$ yr$^{-1}$ (which
according to Kurylov and Kamionkowski \cite{Kurylov:2003ra} is the
limit set by super-Kamiokande).

As the figure shows, the discrete annihilation line is very unlikely
to be observed, even with the next generation instruments. It is just
about detectable for the most promising targets under the most
optimistic assumptions -- the Sagittarius or the Canis Major dSph
galaxies assuming a Moore profile and using next generation ACTs.
Other possible models (such as NFW or cored profiles) and targets
(such as the Galactic Center) are much less propitious still.  For
GLAST, only one line is shown -- namely that for the Canis Major dSph,
but even this lies above all physical mSUGRA models and so provides no
constraints. In particular, monochromatic lines from the Galactic
Center are not visible to GLAST. The difference between this work and
that of \cite{Bergstrom98} is that the latter authors took a very high
dark matter concentration in the center (the profile is just NFW, but
the constant in front is set to ensure maximal flux given two weak
constraints on the mass and the rotation curve). This causes the
$\gamma$-ray flux in monochromatic lines from the Galactic Center as
computed by \cite{Bergstrom98} to be over two orders of magnitude
greater than the values obtained in this paper.

\subsection{Continuum Emission}

\begin{figure}[t]
\includegraphics[height=9cm]{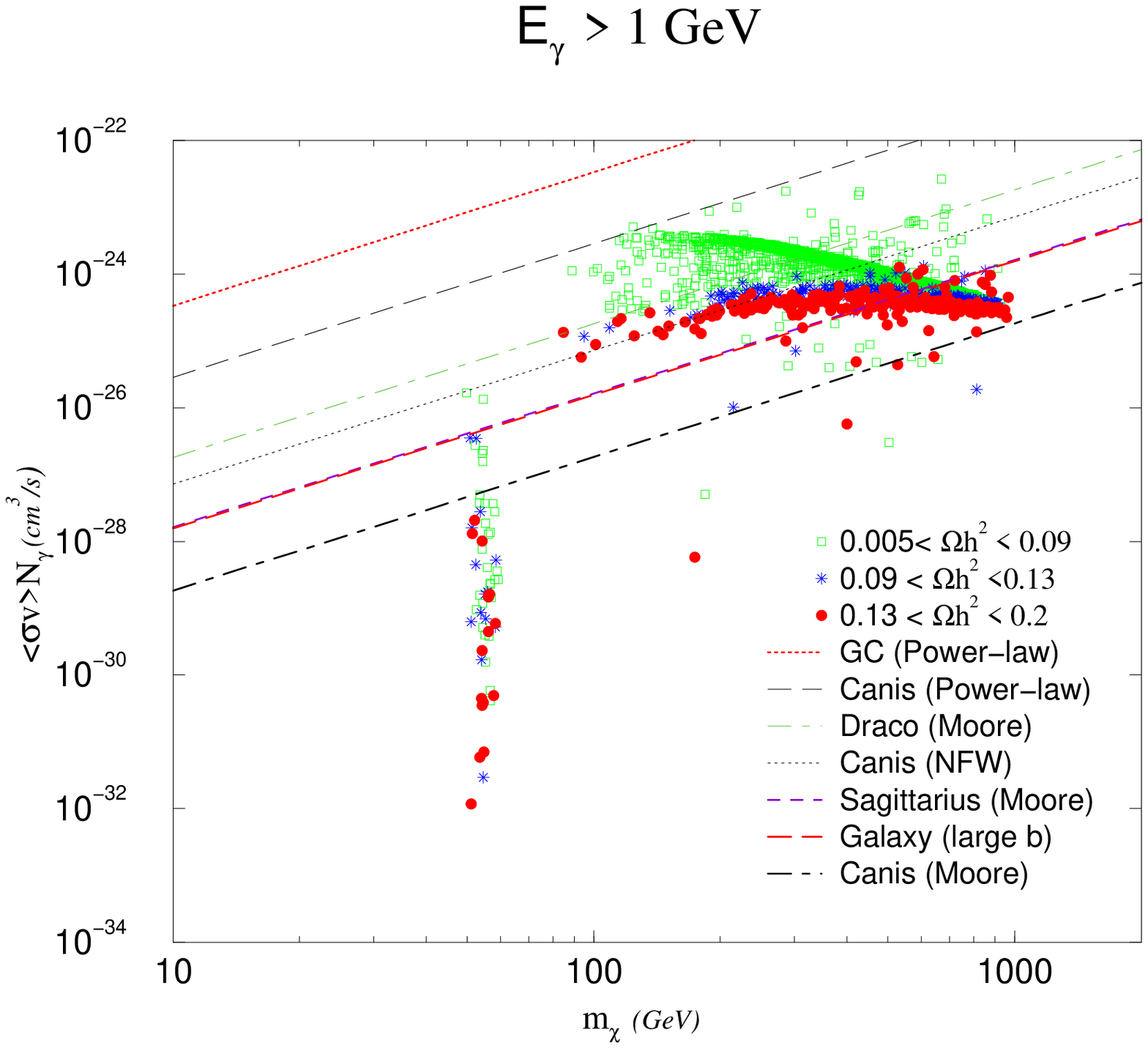}
\includegraphics[height=9cm]{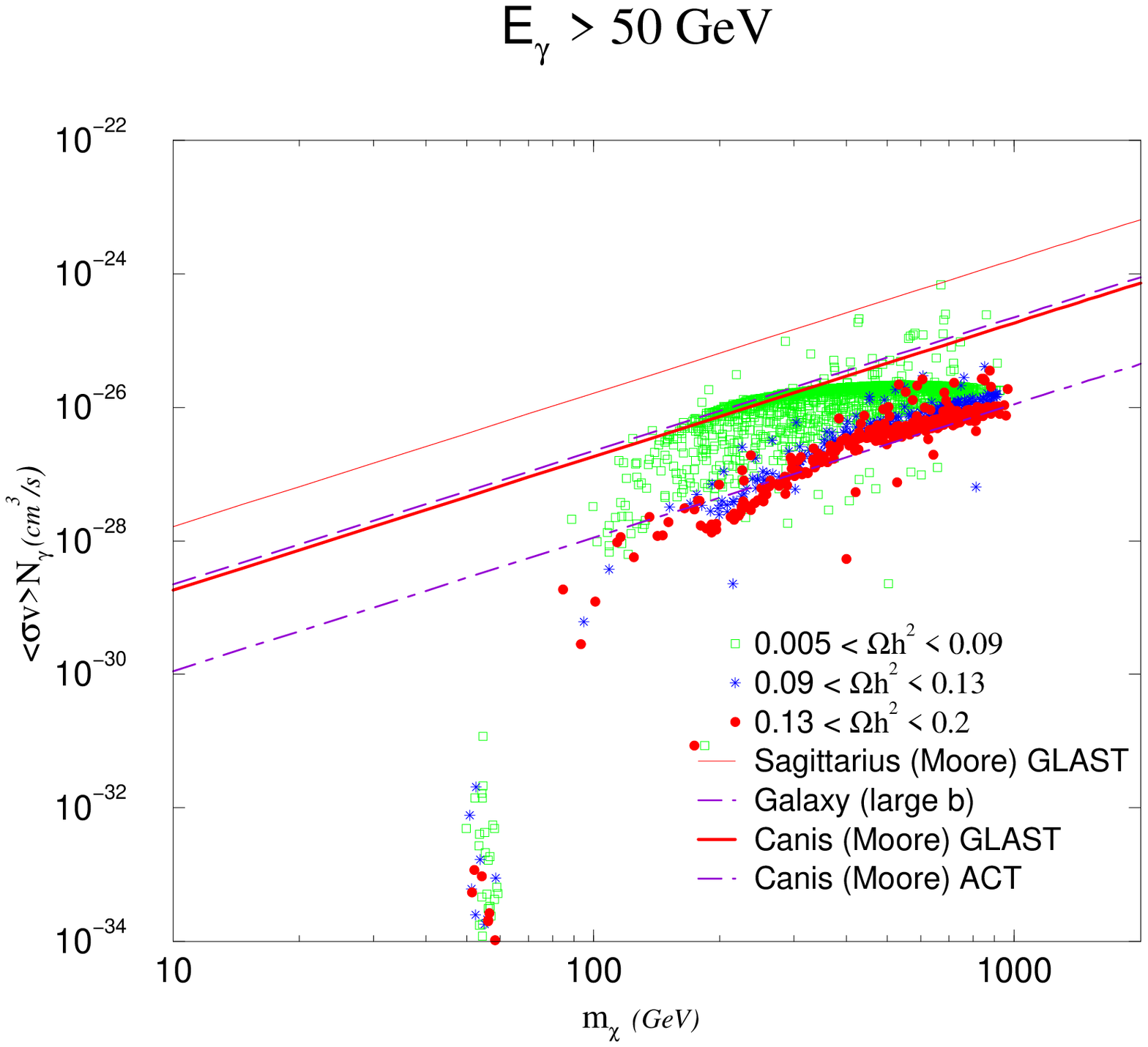}
\caption{Exclusion limits for continuum $\gamma$-ray emission above 1
GeV (top) and 50 GeV (bottom). Only the most favorable cases are
shown.  For $E_\gamma > 1$ GeV, only curves for GLAST are drawn, as
ACTs are insensitive at such low energies. Above 50 GeV, curves are
shown for both GLAST and second generation ACTs.  The green, red and
blue points correspond to mSUGRA models with $\Omega_{\rm CDM} h^2$ in
the range 0.005--0.2, as discussed in Section IIIA. The red points
satisfy the more stringent WMAP constraints $0.09 < \Omega_{\rm CDM}
h^2 < 0.13$.}
\label{fig:continuum}
\end{figure}

The continuum emission comes from hadronization and subsequent pion
decay. The programme DarkSusy \cite{ds} uses results from the PYTHIA
code \cite{pythia} to compute the photon multiplicity for each
neutralino annihilation. Experimental sensitivities are shown in
Fig.~\ref{fig:continuum} for continuum emission above 1 GeV and 50
GeV.

The continuum emission above 1 GeV can yield some
constraints. Although we have computed the curves for four targets
(Draco, Sagittarius, Canis Major and the Galactic Centre) and for the
full range of models in Section II, we give only the most promising
results in the figures. The Draco, Sagittarius and Canis Major dSphs
may yield interesting constraints -- but only if their dark halo
profiles are strongly cusped (the Moore and the NFW profiles both rule
out some supersymmetric models). Unlike the case of the Milky Way,
cusped profiles are still possible for the dSphs. Notice, however,
that substituting cored power-law models for NFW or Moore profiles
causes the exclusion limit to move well above the supersymmetric
parameter space of interest. For $E_\gamma > 1$ GeV, only curves for
GLAST are drawn, as ACTs are insensitive at such low energies.

Also shown in the upper panel of Fig.~\ref{fig:continuum} is a line
corresponding to the Milky Way observed at medium latitudes with the
wide field of view of GLAST, as first suggested by Stoehr et
al. \cite{stoehr}. (This line lies almost exactly on top of the line
for the Sagittarius dSph in the upper panel). Here, the Galaxy has
been modelled with an isothermal power-law model, as opposed to the
cusped models preferred by Stoehr {\it et al.} We agree therefore with
the suggestion of Stoehr {\it et al.} that this is a promising target,
as {\em irrespective of whether the Galaxy is cusped or cored}, there
are always useful constraints on the supersymmetric
parameters. Unfortunately, this attractive option is only available to
GLAST and not for ACTs.

The lower panel of Fig.~\ref{fig:continuum} shows the prospects for
detection of continuum emission above 50 GeV. For ACTs, the Canis dSph
is the best target, though a detectable signal will again only be
measured if the density profile is strongly cusped. For GLAST, the
Galaxy at medium latitudes again leads to some constraints, though not
as strong as when continuum emission above 50 GeV is studied.

\section{Conclusions}

If the dark matter present in the Universe is composed at least in
part by the lightest supersymmetric particle, then this could manifest
itself via $\gamma$-ray emission from pair annihilations.  It is
clearly important to estimate the likely magnitude of the neutralino
annihilation signal. It is also important to identify the likely
locations and spectral r\'egimes in which the signal should be
sought. This paper has provided new estimates of the signal towards
the Galactic Center and the nearby dwarf spheroidals using a variety
of models.

There have been a number of recent calculations predicting that the
neutralino annihilation flux from the inner Galaxy will be detectable
with forthcoming satellites like GLAST and with second generation
atmospheric Cerenkov telescopes (ACTs)
\cite{Bergstrom98,stoehr}. These calculations assume that the cusped
Navarro-Frenk-White (NFW) models for the Milky Way halo hold good. This
assumption is in contradiction with a substantial body of
astrophysical evidence about the inner Galaxy
\cite{evansidm,be,debs}. In any case, even if the Milky Way halo was
originally of NFW form, the formation of the disk and bulge will have
reprocessed the primordial dark matter distribution \cite{bgs}. In
contradiction with earlier results, we do not find the prospects of
detecting the annihilation flux from the Galactic Center to be
particularly promising. In particular, the $\gamma$-ray line coming
from the the $\gamma\gamma$ and $Z\gamma$ final states is not
detectable either with second generation ACTs or with the GLAST
satellite. {\em We caution that many of the recent estimates of high
flux are sensitively dependent on the assumptions made regarding the
innermost structure of the dark halo. Even the best numerical
simulations have difficulty in resolving structures on scales less
than 1 kpc, and so the inner profile is always found by
extrapolation.}

The high mass-to-light ratios of the Local Group dwarf spheroidals
(dSphs) makes them attractive targets.  Cusped profiles like NFW are not
presently ruled out for dSphs like Sagittarius or Draco. It may be
that the visible dwarf galaxy lies entirely within the central parts
of a cusped dark matter halo. If so, then the optimum targets are the
Sagittarius and Canis Major dSphs. The detection of monochromatic
lines is still extremely difficult, but the GLAST satellite may detect
excess continuum $\gamma$ ray emission. This is of course a less
distinctive feature than a sharp line. In particular, if the
Sagittarius or Canis Major dSphs have a strongly cusped dark halo
profile ($\rho \sim r^{-1.5}$ or $\rho \sim r^{-1}$), then some
regions of supersymmetric parameter space can be ruled out. Again,
however, this conclusion only holds good if the dark halo profile is
cusped. Using a cored isothermal-like model for the dark halo, even
the Sagittarius and Canis Major dSphs may be invisible to GLAST and
second generation ACTs.

Unlike Bergstrom {\it et al.} \cite{Bergstrom98}, we do not find the
Galactic Center to be a promising location. Partly, this is because we
believe that the Milky Way does {\em not} have a strongly cusped
profile based on the available astrophysical evidence
\cite{evansidm,be,debs}. Partly, this is because Bergstrom et
al. chose a generous overall normalisation anyhow --- they used the
NFW model corresponding to the maximum flux which satisfies two weak
constraints on the mass and the rotation curve. Accordingly, the local
dark matter density is as high as $\sim 0.6$ GeV cm$^{-3}$ in their
model. When the circular velocity curve of such a halo is combined
with that for the disk and bar, then it necessarily violates the
constraint on the Galactic rotation curve in the inner parts. One
important caveat of our results --- however --- is that the possible
effects of a central black hole are not included in our
calculations. Here, we merely note that the observability of any
expected signal depends on the manner in which the black hole grows
\cite{gs,merritt}.

Stoehr {\it et al.} \cite{stoehr} have also recently emphasised that
the $\gamma$-ray emission from the Galactic Center may have been
overestimated by the use of too strongly cusped profiles. They suggest
that the galaxy at moderate latitudes ($|b| > 10^\circ)$ may also be a
good target for detecting the continuum emission (they do not study
the line emission). This is not really an option for ACTs with their
small field of view. However, {\it it is an attractive possibility for
GLAST, as the continuum emission is detectable irrespective of
uncertainties in halo structure.} For ACTs, the best targets remain
the Sagittarius and Canis Major dSphs.

Very recently, the Large Magellanic Cloud (LMC) has been suggested as
another likely target \cite{olinto}. Judging from \cite{vdm}, the
average mass to light ratio of the LMC within 8.9 kpc is only $\sim 3$
(as opposed to $\sim 100$ for the compact dSphs). This is an upper
limit to the central mass to light ratio. In other words, much as in
the Milky Way, dark matter dominates the outer parts of the LMC and is
responsible for the asymptotic flatness of the ratio curve.  However,
the central parts of the LMC are dominated by the luminous bar and
disk. The assumption that the dark halo dominates the gravitational
potential everywhere is therefore not valid. Hence, the procedure used
in \cite{olinto} of fitting the rotation curve to a NFW dark halo is
flawed. The gravitational potential of the gas and stellar disk and
bar simply cannot be ignored in the central regions.

\begin{acknowledgments}
We are grateful to I. de la Calle, M. Martinez, M. Ramage and C. Tyler
for useful communications. We thank Simon White and the anonymous
Referee for their critical comments on the draft version.

\end{acknowledgments}

\end{document}